\documentclass[12pt]{article}
\usepackage{amsfonts,euscript}

\begin{document}

\title{\bf Unification of twistors and Ramond vectors}
\author {A.A.~Zheltukhin
${}^{{a},{b}}$
\\
{\normalsize 
${}^a$ 
Kharkov Institute of Physics and Technology, 61108 Kharkov, Ukraine,}\\
{\normalsize 
${}^{b}$ 
 Fysikum, AlbaNova, University of Stockholm, SE-10691 Stockholm, Sweden}
\\
{\normalsize  SE-10691, AlbaNova, Stockholm, Sweden}
}                                            
\date{}

\maketitle

\begin{abstract}

We generalize the idea of supertwistors and introduce a  new supersymmetric 
object -- the $\theta$-twistor which includes the {\it composite} Ramond vector $\cite{VZ}$ 
well known from the spinning string dynamics. The symmetries of the chiral 
$\theta$-twistor superspace are studied. 
It is shown that the chiral spin structure introduced by the $\theta$-twistor 
breaks the superconformal boost symmetry but  preserves the scale symmetry
and the super-Poincare symmetry. This geometrical effect of breaking correlates with the 
Gross-Wess effect of the conformal boost breaking for bosonic scattering amplitudes.

\end{abstract}
\section{Introduction}

Conformal invariance is one of the guiding principles both in theory
of particles and condenced matter.  The conception of scale invariance
plays a great role in the description of classical and quantum phase
transitions. In field theories scaling appears as a symmetry of
Lagrangians of massles fields with dimensionless couplings or as an
asymptotic high-energy symmetry of scattering amplitudes. The
superconformal symmetry unifies strings in $AdS_{5}\times S^{5}$ space
and the supersymmetric Yang-Mills theory on its boundary
\cite{M}. Moreover, this symmetry is assumed to be a hidden symmetry
of $M$ theory \cite{W}.

One of widespread convictions is that the scale invariance implies the
conformal invariance. However, the previous study of scale invariant
amplitudes of scalar-scalar, scalar-spinor and scalar-photon scatterings
has shown that the spin structures yield obstacles for the  conformal
symmetry realization \cite{GW}.
Since the spin structure is built in the super-Poincare group,
considered as a fundamental symmetry of the space-time, it is
important to study the relationship between superconformal and scale
symmetries encoded in the geometry of supersymmetric spaces.
 The twistor spaces \cite{PR} and their supersymmetric generalizations 
\cite{Fbr}, \cite{Witt1} are important superspaces connected with superstring and
 super Yang-Mills theories \cite{Witt2}, \cite{Nai}, \cite{Shir}, \cite{BC}, 
\cite{VZ}, \cite{STVZ}, \cite{Znul}, \cite{Sieg1}, \cite{GZ}, \cite{BZ}, 
\cite{Ber}, \cite{UZ}. 
The supertwistor space is invariant under the superconformal  symmetry
generalizing the conformal symmetry of  the twistor space. 

Here we analyze the idea of supertwistor and introduce an alternative 
supersymmetric generalization of the Penrose twistor called the 
$\theta$-twistor. A new property of the $\theta$-twistor is the appearance 
of  the composite Ramond vector (or alternatively a Grassmannian spinor) 
as the additional component ot twistor 
instead of the Grassmannian scalar in the case of the supertwistor.
This result follows from a hidden dual symmetry
of the fundamental quadratic form \cite{Fbr} defining the supertwistor
space and transforming  the scalar Grassmannian component of
 the supertwistor into the  ${\it composite}$ Ramond vector.
 The latter is known as the solution  \cite{VZ} of the Dirac constraint for the
 original  Grassmannian Ramond vector \cite{R} in the models of massless spinning
 particles and strings \cite{BM}, \cite{BSDHDZ}, \cite{BSDH}, \cite{DZ}, \cite{C}, \cite{Zsp}.
We study the symmetry properties of the $\theta$-twistor space 
and find that this space has all of the  symmetries of  the supertwistor space with the 
exception of the  superconformal boosts. 
The  breaking of the superconformal symmetry is a consequence of the
 change of the  Grassmannian  {\it scalar} by the Grassmannian {\it vector} 
or alternatively by {\it spinor} which takes into account a fine spin structure ot the 
chiral $\theta$-twistor superspace.
 This  superspace mechanism of the (super)conformal symmetry breaking
 correlates with the Gross-Wess mechanism of the conformal 
symmetry breaking \cite{GW} just triggered by the substitution of {\it vector} 
or {\it spinor} particles for {\it scalar} particles in their 
 scattering amplitudes. 
  We find that because of the dual symmetry between the $\theta$-twistor and supertwistor 
the invariant Cartan-Volkov differential forms in the supertwistor and $\theta$-twistor spaces 
are also dual. 
These forms may be used for the construction of the dual Wess-Zumino terms and 
invariant actions of particles and strings in the  $\theta$-twistor space.

\section {The supertwistor}

A commuting Weyl spinor  $\nu_{\alpha}$ belonging to the Penrose  spinor doublet 
$(\nu_{\alpha}, \nu^{\beta} x_{\beta\dot\alpha})$  is inert  under the 
transformations  of  $D=4 \,,  N=1$ supersymmetry 
\begin{equation}\label{1}
\begin{array}{c}
\delta\theta_\alpha=\varepsilon_\alpha,\quad 
\delta x_{\alpha\dot\alpha}=
2i(\varepsilon_{\alpha}\bar\theta_{\dot\alpha}-
\theta_{\alpha}\bar\varepsilon_{\dot\alpha}), \quad \delta\nu_{\alpha}=0.
\end{array}
\end{equation}
To introduce the supertwistor \cite{Fbr} we consider the complex superspace 
$(y_{\alpha\dot\alpha},\theta_{\alpha}, \bar\theta_{\dot\alpha})$
and the supersymmetric Cartan-Volkov differential form 
$\omega_{\alpha\dot\alpha}$ associated with the superspace
\begin{equation}\label{2} 
 y_{\alpha\dot\alpha}\equiv x_{\alpha\dot\alpha}-
2i\theta_{\alpha}\bar\theta_{\dot\alpha},\quad
\omega_{\alpha\dot\alpha}=dy_{\alpha\dot\alpha}+ 
4id\theta_{\alpha}\bar\theta_{\dot\alpha}. 
\end{equation}
The invariant scalar form $(\nu\omega\bar\nu)$ constructed from 
$\omega_{\alpha\dot\alpha}$  (\ref{2}) and $\nu_{\alpha},
\bar\nu_{\dot\alpha}$ may be presented as a supersymmetric 
differential form formed by the triplet $Z_{\cal A}$ and its complex 
 conjugate $\bar Z^{\cal A}$
\begin{equation}\label{3}
(\nu\omega\bar\nu)=s(Z,d\bar Z)=-iZ_{\cal A}d\bar Z^{\cal A}.
\end{equation}
The triplets unify the spinors $(\nu^{\alpha},\bar\nu_{\dot\alpha})$ with the
 composite coordinates $q_{\alpha}, \bar q_{\dot\alpha}, \eta, \bar\eta$  
\begin{equation}\label{4}
\begin{array}{c}
Z_{\cal A}\equiv(-iq_{\alpha},\bar\nu^{\dot\alpha}, 2\bar\eta),
\quad
\bar Z^{\cal A}\equiv(\nu^{\alpha},i\bar q_{\dot\alpha}, 2\eta),
\\[0.2cm] 
\eta \equiv \nu^{\alpha}\theta_{\alpha},\quad 
\bar q_{\dot\alpha}= (q_{\alpha})^*\equiv \nu^{\alpha}y_{\alpha\dot\alpha}
=\nu^{\alpha}x_{\alpha\dot\alpha}- 2i\eta\bar\theta_{\dot\alpha}.
\end{array}
\end{equation}
The triplet components form a linear representation of the supersymmetry
\begin{equation}\label{4'}
{\delta \bar q}_{\dot\alpha}=-4i\eta\bar\varepsilon_{\dot\alpha}, \quad
\delta\eta=\nu^{\alpha}\varepsilon_{\alpha}, \quad
\delta\nu_{\alpha}=0.
\end{equation}
 The complex pair $(Z_{\cal A},\bar Z^{\cal A})$ defines the
 supertwistor introduced in \cite{Fbr} as a supersymmetric
 generalization of the projective Penrose twistor.

 The  supertwistor space may be equivalently defined as a complex projective
 superspace equipped with the invariant bilinear null form $s(Z, \bar Z')$ 
\begin{equation}\label{5}
\begin{array}{c}
s(Z, \bar Z')\equiv -iZ_{\cal A}\bar Z'^{\cal A}= -q_{\alpha}\nu'^{\alpha}+ 
\bar\nu^{\dot\alpha}\bar q'_{\dot\alpha} -4i\bar\eta\eta'=0,
\end{array}
\end{equation}
 where the complex conjugate triplet $\bar Z'^{\cal A}$ is given by 
 (\ref{4}) with $\nu'$ substituted for $\nu$
\begin{equation}\label{6}
\begin{array}{c}
\bar Z'^{\cal A}\equiv (\nu'^{\alpha},i\bar q'_{\dot\alpha}, 2\eta'),
\quad 
\bar q'_{\dot\alpha}=\nu'^{\alpha}y_{\alpha\dot\alpha}, \quad \eta'=
\nu'^{\alpha}\theta_{\alpha}.
\end{array}
\end{equation}
 The quadratic form (\ref{5}) is invariant under the global superconformal 
symmetry as it was shown in \cite{Fbr}. The fermionic sector of the supertwistor
 (\ref{4}) contains only the {\it scalar}  projection $\eta$  of  $\theta_{\alpha}$. 
It sets the question: whether it is possible to preserve the superconformal 
symmetry without the reduction of the $\theta$ components while the twistor supersymmetrization?

\section{ The $\theta$-twistor: A unification of Penrose twistor and Ramond vector}

A characteristic point in the supertwistor construction is the unification of the 
chiral coordinates $(y_{\alpha\dot\alpha},\theta_{\alpha})$ with the Penrose
spinor $\nu_{\alpha}$ having the ${\it same}$ chirality as
$\theta_{\alpha}$. 
The chirality ${\it coincidence}$ permits to
construct the  ${\it left\, projection}$ of the coordinates 
 $(y_{\alpha\dot\alpha},\theta_{\alpha})$ on the spinor  $\nu^{\alpha}$ transforming the complex vector 
$y_{\alpha\dot\alpha}$ into the spinor 
$\bar q_{\dot\alpha}=
\nu^{\alpha}
y_{\alpha\dot\alpha}$  
and the spinor 
$\theta_{\alpha}$ into the complex scalar $\eta=\nu^{\alpha}\theta_{\alpha}$,
i.e. making a step down in  spins:
$(1,{\frac{1}{2}})\rightarrow (\frac{1}{2},0)$.
This projection preserves the ${\it linear}$ character of the supersymmetry transformations in the chiral
 space \begin{equation}\label{4''}
\delta y_{\alpha\dot\alpha}= -4i\theta_{\alpha}\bar\varepsilon_{\dot\alpha},\,\quad \delta\theta_\alpha=
\varepsilon_\alpha
\end{equation}
 and yields the linear realization (\ref{4'}) of the supersymmetry by the triplet $\bar Z^{\cal A}$ (\ref{4}).

An alternative supersymmetric  triplet was proposed in \cite{Z}, where 
 the chiral coordinates
$(y_{\alpha\dot\alpha},\theta_{\alpha})$ were  unified  with the
c.c. Penrose spinor $\bar\nu_{\dot\alpha}$ whose chirality has ${\it
opposite}$ sign to the $\theta_{\alpha}$ chirality. 
In that case one can not construct a projection of $\theta_{\alpha}$ on  $\bar\nu_{\dot\alpha}$
  reducing the number of the fermionic  coordinates. However,   
in the bosonic sector one can construct the new composite spinor $l_{\alpha}$
 produced by the ${\it right\, projection}$
 of the ${\it chiral}$ coordinate $y_{\alpha\dot\alpha}$ on $\bar\nu^{\dot\alpha}$
\begin{equation}\label{7}
l_{\alpha}\equiv y_{\alpha\dot\alpha}\bar\nu^{\dot\alpha}
=x_{\alpha\dot\alpha}\bar\nu^{\dot\alpha}- 2i\theta_{\alpha}\bar\eta, \quad 
l_{\alpha}=q_{\alpha}- 4i\theta_{\alpha}\bar\eta.
\end{equation}
Because the complex matrix $y_{\alpha\dot\alpha}$ is not a Hermitian its right 
and left contractions with spinors are not connected by complex conjugation   
\begin{equation}\label{ql}
(\bar q_{\dot\alpha})^* =  l_{\alpha}+ 4i\theta_{\alpha}\bar\eta, 
\quad
\nu^{\alpha}l_{\alpha}=\bar q_{\dot\alpha}\bar\nu^{\dot\alpha}.
\end{equation}
Fixation of  $\theta_{\alpha}$  as a superpartner of $l_{\alpha}$ results in 
  the ${\it nonlinear}$ realization of the supersymmetry (\ref{1}) by the  ${\it three \, spinors}$ 
\begin{equation}\label{8}
{\delta l}_{\alpha}=-4i\theta_{\alpha}(\bar\nu^{\dot\beta}
\bar\varepsilon_{\dot\beta}), \quad
\delta\theta_{\alpha}=\varepsilon_{\alpha}, \quad
\delta\bar\nu_{\dot\alpha}=0
\end{equation}
which form the new complex spinor  triplet $\Xi_{\cal A}$ \cite {Z}
\begin{equation}\label{9}
\Xi_{\cal A}\equiv(-il_{\alpha},\bar\nu^{\dot\alpha},\theta^{\alpha}),\quad
\bar\Xi^{\cal A}\equiv(\Xi_{\cal A})^*
=(\nu^{\alpha},i{\bar l}_{\dot\alpha},\bar\theta^{\dot\alpha}).
\end{equation}
 In the $\Xi$-triplet space the supersymmetry generators take the form 
\begin{equation}\label{48/1}
\begin{array}{c}
Q^{\alpha}=\frac{\partial}{\partial\theta_{\alpha}}+
4i\nu^{\alpha}(\bar\theta_{\dot\beta}\frac{\partial}{\partial\bar l_{\dot\beta}}),
 \quad
 \bar Q^{\dot\alpha}\equiv -(Q^{\alpha})^*=
\frac{\partial}{\partial\bar\theta_{\dot\alpha}}+ 
4i\bar\nu^{\dot\alpha}(\theta_{\beta}\frac{\partial}{\partial l_{\beta}})
\end{array}
\end{equation}
with their anticommutator closed by the vector generator $ P^{\dot\beta\alpha}=
(\bar\nu^{\dot\beta}\frac{\partial}{\partial l_{\alpha}}+
\nu^{\alpha}\frac{\partial}{\partial\bar l_{\dot\beta}})$ 
\begin{equation}\label{49/1}
\begin{array}{c}
\{ Q^{\alpha}, \bar Q^{\dot\beta}\}= 4iP^{\dot\beta\alpha},\, \quad
[ Q^{\gamma},P^{\dot\beta\alpha}]=[\bar Q^{\dot\gamma},P^{\dot\beta\alpha}]= 
\{ Q^{\gamma},Q^{\beta}\}= \{\bar Q^{\dot\gamma}, \bar Q^{\dot\beta}\}=0.    
\end{array}
\end{equation}
The quadratic form (\ref{5}) expressed in terms of $\Xi_{\cal A}$ and 
$\bar\Xi^{\cal A}$ (\ref{9}) becomes a ${\it nonlinear}$ form
\begin{equation}\label{10}
\begin{array}{c}
s(Z,\bar Z')\equiv
-iZ_{\cal A}\bar Z'^{\cal A}=\tilde s(\Xi,\bar\Xi')\equiv 
 -l_{\alpha}\nu'^{\alpha}+ 
\bar\nu^{\dot\alpha}\bar l'_{\dot\alpha} - 
4i(\nu'_{\alpha}\bar\nu_{\dot\alpha})\theta^{\alpha}\bar\theta^{\dot\alpha}=0 
\end{array}
\end{equation} 
 in the $\Xi$-triplet space. 
That nonlinearity is a consequence of the fixation of $\theta_{\alpha}$ as the superpartner of
 $l_{\alpha}$. 
However, this fixing is not the only possible. 
 In fact, to construct  the supertwistor $Z$-triple the spin ${down}$ transition
 $(1,{\frac{1}{2}})\rightarrow (\frac{1}{2},0)$ was used. 
But,
there is an alternative way described by the spin transition
 $(1,{\frac{1}{2}})\rightarrow (\frac{1}{2},1)$. 
This way proposes the complex vector $(\theta_{\alpha}\bar\nu_{\dot\beta})$ as a superpartner of the spinor 
$l_{\beta}$.
With  this observation we multiply the second equation in (\ref{8}) by $\bar\nu^{\dot\beta}$ and  obtain 
 Eqs. (\ref{8}) just rewritten in the  desired linear form
\begin{equation}\label{8R}
{\delta l}_{\alpha}=4i(\theta_{\alpha}\bar\nu_{\dot\beta})
\bar\varepsilon^{\dot\beta}, \quad
\delta(\theta_{\alpha}\bar\nu_{\dot\beta})=\varepsilon_{\alpha}\bar\nu_{\dot\beta}, \quad
\delta\bar\nu_{\dot\alpha}=0,
\end{equation}
or equivalently in the explicit linear form as
\begin{equation}\label{8'}
{\delta l}_{\alpha}=-4i(\sigma_{m}\bar\varepsilon)_{\alpha}\bar\eta^{m}, \quad
\delta\bar\eta_{m}=-\frac{1}{2}(\varepsilon\sigma_{m}\bar\nu), \quad
\delta\bar\nu_{\dot\alpha}=0.
\end{equation} 
The grassmannian vectors $\bar\eta_{m}$ and $\eta_{m}$  in (\ref{8'}) are the {\it composite} Ramond vectors   
\begin{equation}\label{vz}
\begin{array}{c} 
\eta_{m}\equiv -\frac{1}{2}(\nu\sigma_{m}\bar\theta),\quad
\bar\eta_{m}=(\eta_{m})^{*}= -\frac{1}{2}(\theta\sigma_{m}\bar\nu),
\\[0.2cm]
\nu_{\beta}\bar\theta_{\dot\alpha}\equiv \eta_{\beta\dot\alpha}=
(\sigma^{m})_{\beta\dot\alpha}\eta_{m},\quad  \eta_{m}\eta_{n}+\eta_{n}\eta_{m}=0.
\end{array}
\end{equation} 
previously introduced in \cite{VZ} (see details in \cite{Z2}) 
to prove the equivalence between superparticles and spinning particles.
In terms of the Ramond vectors the nonlinear term $ 4i(\nu'_{\alpha}\bar\nu_{\dot\alpha})
\theta^{\alpha}\bar\theta^{\dot\alpha}$ in (\ref{10}) is presented in the {\it bilinear} form 
\begin{equation}\label{nc}
\begin{array}{c}
 -4i(\nu'_{\alpha}\bar\nu_{\dot\alpha})\theta^{\alpha}\bar\theta^{\dot\alpha}
\equiv
4i\bar\eta\eta'\equiv2i(\bar\nu\tilde\sigma_{m}\theta)(\nu'\sigma^{m}\bar\theta)
\equiv-8i\bar\eta_{m}\eta'_{m}.
\end{array}
\end{equation}
As a result, the quadratic  form (\ref{10}) defining the supertwistor space
 becomes the new {\it quadratic}  form 
\begin{equation}\label{10'}
\begin{array}{c} 
s=s(Z,\bar Z')=s{(\bf \Xi,\bf \bar\Xi')}
\equiv-i{\bf\Xi_{\cal A}}
{\bf\bar\Xi'^{\cal A}}= 
-l_{\alpha}\nu'^{\alpha}+ 
\bar\nu^{\dot\alpha}{\bar l'}_{\dot\alpha} - 8i\bar\eta_{m}\eta'^{m}=0
\end{array}
\end{equation}
in the complex projective space of the $\bf\Xi$-triplets including  the composite
 Ramond vector 
\begin{equation}\label{sqrf}
\begin{array}{c} 
{\bf\Xi_{\cal A}}
\equiv(-il_{\alpha},\bar\nu^{\dot\alpha},2\sqrt{2}\bar\eta_{m}),\quad
{\bf\bar\Xi^{\cal A}}\equiv({\bf\Xi}_{\cal A})^*=
(\nu^{\alpha},i{\bar l}_{\dot\alpha},2\sqrt{2}\eta^{m}).
\end{array}
\end{equation}
So, the supertwistor $Z$-triplet (\ref{4})  transforms into the $\bf\Xi$- triplet 
(\ref{sqrf}) under the substitution $(q_{\alpha}, \bar\eta) \rightarrow 
(l_{\alpha}, \sqrt{2}\bar\eta_{m})$ producing the new quadratic representations of 
the quadratic form (\ref{5}) but now in terms of the $\bf \Xi$-triplet.  
Thus,  the $Z$-triplet (\ref{4}) and the $\bf \Xi$-triplet (\ref{sqrf}) occur to be equal in 
their own rights and it gives  a reason to call the new 
$\bf \Xi$-triplet (\ref{sqrf})
(or equivalently the $\Xi$-triplet (\ref{9})) the $\theta$-twistor to emphasize
 its difference from the  supertwistor. 
 
The  $\theta$-twistor  independently appears as a geometrical object alternative to the
 supertwistor
from another point of view. The latter bases on the observation that 
the supertwistor and the $\theta$-twistor are the general solutions of 
${\it different}$ supersymmetric constraints. 
The supertwistor solves the
generalized chiral constraint in the superspace 
$(y_{\alpha\dot\alpha},\,\theta_{\alpha})$ complemented by the 
{\it left} Weyl spinor $\nu_{\alpha}$ and the new scalar operator
$\nu_{\alpha}D^{\alpha}$, 
\begin{equation}\label{61/1}
\begin{array}{c} 
{\bar D}^{\dot\alpha}F(x,\theta,\bar\theta)=0
\longrightarrow \, F=F(y,\theta), \\[0.2cm]
\nu_{\alpha}D^{\alpha}F(y,\theta,\nu)=0 \longrightarrow \, F=F(\bar
Z^{\cal A}).  \end{array}, 
\end{equation} 
where $F(\bar Z^{\cal A})$ is the superfield depending on the triplet $\bar Z^{\cal A}$.
  In contrast, the $\theta$-twistor solves the supersymmetric constraints
in the chiral space complemented by the {\it right} Weyl spinor
$\bar\nu_{\dot\alpha}$ and the Dirac chiral operator
$\bar\nu_{\dot\alpha}\frac{\partial}{\partial x_{\alpha\dot\alpha}}$,
\begin{equation}\label{64/1} 
\begin{array}{c} {\bar
D}^{\dot\alpha}F(x,\theta,\bar\theta)=0 \longrightarrow \,
F=F(y,\theta), \\[0.2cm] \bar\nu_{\dot\alpha}\frac{\partial}{\partial
x_{\alpha\dot\alpha}}F(y,\theta, \bar\nu)=0 \longrightarrow \,
F=F(\Xi_{\cal A}).  
\end{array} \end{equation} 
It is easy to see that
the additional Dirac constraint in (\ref{64/1}), selecting the $\Xi $-triplet,
 may be rewritten in an
 equivalent form using the composite  Ramond  vector. 
It follows from the multiplication of  the second of  Eqs. (\ref{64/1}) by  $\theta_{\alpha}$ 
\begin{equation}\label{Dir}
\begin{array}{c}
\theta_{\alpha}\bar\nu_{\dot\alpha}\frac{\partial}{\partial
x_{\alpha\dot\alpha}}F(y,\theta, \bar\nu) \equiv
\bar\eta_{\alpha\dot\alpha}\frac{\partial}{\partial
x_{\alpha\dot\alpha}}F(y,\theta,
\bar\nu)\equiv\bar\eta^{m}\partial_{m}F(y,\theta,\bar\nu) =0.  
\end{array}
\end{equation}
 
We see that the $\theta$-twistor is an object dual to the supertwistor but 
 preserving all of the spin degrees of freedom encoded by the spinor $\theta_{\alpha}$. 
Then our question  about the superconformal symmetry  transforms to the question
 whether all of the symmetries of supertwistor survive the transition to the $\theta$-twistor? 
In other words what is the payment for the restoration of the spin degrees  of freedom lost 
by the supertwistor?  Studying the $\theta$-twistor symmetries is necessary to answer the question. 

It is easy to see the invariance of the nonlinear and  quadratic representations 
 (\ref{10'}), (\ref{10}) under the supersymmetry (\ref{8}), (\ref{8'}), 
the scale and phase transformations  
\begin{equation}\label{52/2}
\begin{array}{c}
l'_{\beta}=e^{\varphi}l_{\beta}, 
\quad
{\bar l}'_{\dot\beta}=e^{\varphi*}{\bar l}_{\dot\beta},
 \quad
\nu'_{\beta}=e^{-\varphi}\nu_{\beta} ,
\quad 
\bar\nu'_{\dot\beta}=e^{-\varphi*}\bar\nu_{\dot\beta},
\quad
\theta'_{\beta}=e^{\varphi}\theta_{\beta}, 
\quad
\bar\theta'_{\dot\beta}=e^{\varphi*}\bar\theta_{\dot\beta},
\end{array}
\end{equation}
 described  by the complex parameter $\varphi=\varphi_{R}+i\varphi_{I}$, 
as well as under the $\gamma_5$ rotations  \cite{Z}
 \begin{equation}\label{50}
 \theta'_{\beta}=e^{i\lambda}\theta_{\beta}, \quad 
\bar\theta'_{\dot\beta}=e^{-i\lambda}\bar\theta_{\dot\beta} \,.
\end{equation}
 So, the  $\theta$-twistor triplet is  closed under these symmetries forming
 their representations similarly the $Z$-triplet. But $Z$-triplet  also realizes
 the (super)conformal boosts \cite{WZ} and forms a  representation of the superconformal group.
What is about the closure of the $\Xi$ and  $\bf\Xi$ triplets under the superconformal 
boosts $S^{\alpha},\,\bar S^{\dot\alpha}$? We discuss the question below.

\section{$\theta$-twistor and superconformal symmetry breaking}

To answer the  question whether the $\theta$-twistor forms a representation of the superconformal boosts, 
let us remind the superboost realization \cite{Z} in  the chiral superspace $(y_{\alpha\dot\beta}, \theta_{\alpha})$  
\cite{WB}
\begin{equation}\label{36/1} 
\delta y_{\alpha\dot\alpha}=4i\theta_{\alpha}(\xi^{\beta}y_{\beta\dot\alpha}),
\quad
 \delta\theta_{\alpha}=- y_{\alpha\dot\beta}\bar\xi^{\dot\beta} +
 4i\theta_{\alpha}(\xi^{\beta}\theta_{\beta}). 
\end{equation} 
The left multiplication of (\ref{36/1}) by $\nu^{\alpha}$ yields  the variations of the supertwistor 
components 
 \begin{equation}\label{36/1'} 
 \delta\bar q_{\dot\alpha}=(\delta\nu^{\beta}+ 4i\eta\xi^{\beta})y_{\beta\dot\alpha}
,\quad
\delta\eta= (\delta\nu^{\beta}+ 4i\eta\xi^{\beta})\theta_{\beta} + (\bar q^{\dot\alpha}\bar\xi_{\dot\alpha}), 
\end{equation}
 where dependence on $y$ and $\theta$ is eliminated  by the choice of the transformation law for $\nu^{\beta}$ 
 \begin{equation}\label{36/1''} 
\delta\nu^{\beta}=- 4i\eta\xi^{\beta}, \quad   \delta\eta= (\bar q^{\dot\alpha}\bar\xi_{\dot\alpha}),
\quad  
\delta\bar q_{\dot\beta}=0.
\end{equation} 
Eqs. $(\ref{36/1''})$  define the variations of the supertwistor under the superconformal boosts.

A similar procedure, realized by multiplication of $\delta y_{\alpha\dot\alpha}$  from (\ref{36/1})
by $\bar\nu^{\dot\alpha}$, yields the superconformal boost transformations of $l_{\alpha}$  but 
with  the 
 {\it right} index of $y_{\alpha\dot\beta}$ occupied by  $\delta\bar\nu^{\dot\beta}$
\begin{equation}\label{36/1t}  
 \delta l_{\alpha}=  y_{\alpha\dot\beta}\delta\bar\nu^{\dot\beta} + 4i\theta_{\alpha}(\xi^{\beta}l_{\beta}),
\quad
 \delta\theta_{\alpha}=- y_{\alpha\dot\beta}\bar\xi^{\dot\beta} +  4i\theta_{\alpha}(\xi^{\beta}\theta_{\beta}).
\end{equation}
In the variation of $\theta_{\alpha}$ (\ref{36/1t}) we also observe the {\it right} 
index of $y_{\alpha\dot\beta}$ 
contracted with the parameter $ \bar\xi^{\dot\beta}$ index.
It means that 
 the minimal complex dimension of the bosonic superpartner 
of $\theta_{\alpha}$ needed for the superconformal boosts realization equals four  and  it 
 prevents a reduction of $y_{\alpha\dot\beta}$ into $l_{\alpha}$.
In the supertwistor case the transmutation of $y_{\alpha\dot\beta}$ into  
$\bar q_{\dot\beta}$ has been accompanied by a synchronous transmutation of $\theta_{\alpha}$ 
into the grassmannian {\it scalar} $\eta$ whose component number is also half of the component number of 
the {\it spinor} $\theta_{\alpha}$.
Because of this double reduction there was not breaking in the superconformal symmetry realization by the 
$\bar Z$-triplet
 forming the antiholomorphic sector of the supertwistor space. 
On the  contrary, in the case of $\Xi$-triplets we preserve the spinor
$\theta_{\alpha}$  from a reduction but transmute the complex vector $y_{\alpha\dot\beta}$ into the spinor 
$l_{\alpha}$ 
whose number of components equals half of the component number of $y_{\alpha\dot\beta}$. 
The reduction yields a {\it deficit} in  the bosonic sector of $\Xi_{\cal A}$. 
To compensate  the deficit the components of  $\Xi_{\cal A}$
 must be extended at least by one auxiliary spinor. 

To check  this observation we shall treat $\nu_{\alpha}$ as one of 
the Newman-Penrose basis elements in the spinor space \cite{PR} and add an auxiliary spinor 
$v_{\alpha}$, 
defined by the relations 
\begin{equation}\label{np} 
 \nu^{\alpha}v_{\alpha}=1, \quad  v_{\alpha}\nu^{\beta}-  \nu_{\alpha}v^{\beta} = \delta^{\beta}_{\alpha},
\end{equation}
which may be used  as the second element of the spinor dyad.
Taking into account the completness condition from Eqs. (\ref{np}) we present the variation 
of $\theta_{\alpha}$ (\ref{36/1t}) in the form
\begin{equation}\label{exp1}
\delta\theta_{\alpha}=-l_{\alpha}\bar\xi + m_{\alpha}\bar\zeta + 4i\theta_{\alpha}(\xi^{\beta}\theta_{\beta}),
\end{equation}
where $\bar\xi,\,\bar\zeta$ 
are  effective superboost parameters given by the projections of $\bar\xi^{\dot\beta}$ on the dyad
\begin{equation}\label{m0}
  \xi=(\xi^{\beta}v_{\beta}),\quad \zeta=(\xi^{\beta}\nu_{\beta}),  \quad   m_{\alpha}= (y\bar v)_{\alpha}
\end{equation}
and the new  independent spinor $m_{\alpha}$ is the projection of $y_{\alpha\dot\beta}$ on the spinor $\bar v^{\dot\beta}$.  
The superboost of $l_{\alpha}$ (\ref{36/1t}) can also be  presented by a similar expansion in the 
spinors $l_{\alpha},\,
m_{\alpha}$ and $\theta_{\alpha}$
\begin{equation}\label{36/0} 
\delta l_{\alpha}=l_{\alpha} 
(\bar v_{\dot\alpha}\delta\bar\nu^{\dot\alpha}) -
m_{\alpha} (\bar\nu_{\dot\alpha}\delta\bar\nu^{\dot\alpha}) +
4i\theta_{\alpha}(\xi^{\beta}l_{\beta}).
 \end{equation}
The expansions (\ref{exp1}), (\ref{36/0}) confirm  the  need to extend the holomorphic $\Xi_{\cal A}$ triplet
by the auxiliary spinor $m_{\alpha}$ to realize the superconformal boosts.

Otherwise, it  might be  that  the $\Xi_{\cal A}$-triplet could  realize a  ${\it half}$ of the superconformal 
boosts.  It  requires the absence of  $m_{\alpha}$ in Eqs. (\ref{exp1}), (\ref{36/0}) that is equivalent 
to the conditions 
\begin{equation}\label{abse}
\bar\zeta=(\bar\xi^{\dot\alpha}\bar\nu_{\dot\alpha})=0,  
\quad (\bar\nu_{\dot\alpha}\delta\bar\nu^{\dot\alpha})=0 
\end{equation} 
whose solutions contain the spinor $\nu_{\alpha}$  belonging to the $\bar\Xi^{\cal A}$-triplet 
from the ${\it antiholomorphic}$ sector of the $\theta$-twistor space
\begin{equation}\label{red/0}
 \quad  \xi_{\alpha}=\nu_{\alpha}\xi, \quad
\delta\nu_{\alpha}=\nu_{\alpha}\delta\varphi
\end{equation} 
and consequently introduces the auxiliary spinor  $\nu_{\alpha}$ instead of  $m_{\alpha}$.
It proves that the  holomorphic $\Xi_{\cal A}$ triplet does not realize even the half of the superboosts parametrized by $\xi$.

However, the solution (\ref{red/0}) points out on a possibility  to realize the half of superconformal boosts 
as a symmetry of the {\it whole} $\theta$-twistor space which {\it mixes} its holomorphic $\Xi_{\cal A}$ 
and  antiholomorphic  $\bar\Xi^{\cal A}$ triplets. The  substitution of  (\ref{red/0})  in  (\ref{36/0}) 
 yields the realization of the
half of superboosts in the form   
\begin{equation}\label{vari}
\begin{array}{c}
\delta \theta_{\alpha}=-l_{\alpha} \bar\xi - 
4i\theta_{\alpha}(\nu^{\beta}\theta_{\beta})\xi \,, 
\quad 
\delta l_{\alpha}=l_{\alpha}\delta\bar\varphi + 
4i\theta_{\alpha}(\nu^{\beta}l_{\beta})\xi, 
\quad \,
\delta\bar\nu^{\dot\alpha}= \bar\nu^{\dot\alpha}\delta\bar\varphi.
\end{array}
\end{equation} 
The variation $\delta\bar\varphi$ is fixed by the  condition 
\begin{equation}\label{red'}
\delta \bar\varphi = 4i(\nu^{\beta}\theta_{\beta})\xi \,\rightarrow  \,
\delta\bar\eta_{m}=
\frac{1}{2}(l\sigma_{m}\bar\nu)\bar\xi.
\end{equation}
which confines  the transformed Ramond vector $\bar\eta_{m}$ inside of the ${\bf\Xi_{\cal A}}$ triplet.
The substitution of $\delta \bar\varphi$ (\ref{red'}) in Eqs. (\ref{vari}) transforms them to the form
\begin{equation}\label{fvar}
\begin{array}{c} 
\delta l_{\alpha}=4i[\, (\nu^{\beta}\theta_{\beta})l_{\alpha}+ 
(\nu^{\beta}l_{\beta})\theta_{\alpha} \,]\xi \,, \quad 
\delta\bar\nu^{\dot\alpha}=4i\bar\nu^{\dot\alpha}(\nu^{\beta}\theta_{\beta})\xi\,,
 \\[0.2cm]
\delta\theta_{\alpha}= - l_{\alpha}\bar\xi - 
4i\theta_{\alpha}(\nu^{\beta}\theta_{\beta})\xi. 
\end{array}
\end{equation} 
 Eqs. (\ref{fvar}) contain the spinor $\nu^{\alpha}$ from the antiholomorphic $\bar\Xi^{\cal A}$ 
triplet and the equations must be added by their complex conjugate to take into account the 
superboost of  $\nu^{\alpha}$. 
It shows  that the whole $\theta$-twistor space 
closes under the half of the superconformal boosts which mix  the ${\it holomorphic}$ and 
${\it antiholomorphic}$ sectors formed by the triplets  $\Xi_{\cal A}$ and  $\bar\Xi^{\cal A}$
although each of them is not closed.  

The problem of the  superconformal boost realization more sharpens for the case of the
$\bf\Xi_{\cal A}$ triplet (\ref{sqrf}) containing the Ramond vector $\bar\eta_{m}$ 
as the superpartner of $l_{\alpha}$ instead of $\theta_{\alpha}$. 

Here we encounter with the problem of transmutation of  
 $\theta_{\alpha}$ into $\bar\eta_{m}$ without an extension  of the $\theta$-twistor space.
 It is  seen from Eqs. (\ref{fvar}) defining the  superboosts of the constituents of the Ramond vector 
$(\theta_{\alpha}\bar\nu_{\dot\beta})$. After multiplication of  
$\delta\theta_{\alpha}$ by $\bar\nu^{\dot\beta}$ (\ref{fvar})
and  taking into account the relation $\nu^{\beta}l_{\beta}=\bar q_{\dot\beta}\bar\nu^{\dot\beta}$ (\ref{ql})
we obtain the superboost realization including the desired Ramond vector $\bar\eta_{m}$
\begin{equation}\label{vect}
\begin{array}{c} 
\delta l_{\alpha}=4i[\, \eta l_{\alpha} - (\sigma^{m}{\bar q})_{\alpha} \bar\eta_{m} \,]\xi \,, 
\quad 
\delta\bar\nu_{\dot\alpha}=4i(\nu\sigma^{m})_{\dot\alpha}\bar\eta_{m} \xi, \quad 
\delta\bar\eta_{m}=\frac{1}{2}(l\sigma_{m}{\bar\nu}) \bar\xi.
 \end{array}
\end{equation}  
In the variation $\delta l_{\alpha}$ (\ref{vect}) we observe the appearance of 
${\bar q}_{\dot\alpha}$ which belongs neither to the $\bf\Xi_{\cal A}$ nor to $\bf\bar\Xi^{\cal A}$ 
triplets but to the supertwistor triplet $\bar Z^{\cal A}$ (\ref{4}). 
 It proves that the whole complex superspace formed by  ${\bf\Xi_{\cal A}}$ and ${\bf\bar\Xi^{\cal A}}$ 
triplets is not closed under
 the supeconformal boosts and each of these triplets form representations of only the maximal subgroup 
of the superconformal group.

The superconformal symmetry breaking is explained by the difference of {\it chiralities} 
of the ${\bar\nu}^{\dot\alpha}$ and $\theta_{\alpha}$ spinors forming  the $\Xi_{\cal A}$ or 
${\bf\Xi_{\cal A}}$ triplets. This effect correlates with the Gross-Wess mechanism 
associated with the nontrivial spin structure of the  scattering amplitudes.

\section {Dual Wess-Zumino terms and dual actions}
 
The $\theta$-twistor introduces an alternative supersymmetric extension of the Penrose twistor 
 and deserves of studying in various physical applications.  
 It was  previously shown that using the chiral superspace
associated  with the $\theta$-twistors  yields   
 an  infinite  chain of  higher spin chiral supermultiplets 
$(\frac{1}{2}, 1), (1, \frac{3}{2}), (\frac{3}{2}, 2),....,(S, S + \frac{1}{2})$ 
generalizing the scalar supermultiplet   \cite{Z}.
 These supermultiplets include the auxiliary $F$-field absent in the supertwistor description. 

Another interesting problem is to study  actions of 
supersymmetric models of particles, strings and  branes in the $\theta$-twistor space. 
The  actions  may be constructed using the supersymmetric 
differential forms in the $\theta$-twistor space. 
It is illustrated by a simple example of the supersymmetric 
one-form (\ref{3}) which  has two dual representations
\begin{equation}\label{3'}
\begin{array}{c} 
(\nu\omega\bar\nu)=
(Z,d\bar Z)=-iZ_{\cal A}d\bar Z^{\cal A}= -q_{\alpha}d\nu^{\alpha}+ 
\bar\nu^{\dot\alpha}d{\bar q}_{\dot\alpha} - 4i{\bar\eta}d\eta\\
[0.2cm]
= s({\bf\Xi},d {\bf\bar\Xi)}=-i{\bf\Xi_{\cal A}}
d{\bf\bar\Xi^{\cal A}}=  -l_{\alpha}d\nu^{\alpha}+ 
\bar\nu^{\dot\alpha}d{\bar l}_{\dot\alpha} - 8i{\bar\eta}_{m}d\eta^{m},
\end{array}
\end{equation}
as a consequence of the above discussed dual symmetry connecting $\theta$-twistor 
and supertwistor. 
The differential forms (\ref{3'}) may be presented  in the equivalent form 
\begin{equation}\label{3L}
\begin{array}{c}
-iZ_{\cal A}d\bar Z^{\cal A}=
[\,\nu^{\alpha}dq_{\alpha}+ 
\bar\nu^{\dot\alpha}d\bar q_{\dot\alpha} - 
2i(\bar\eta d\eta- d\bar\eta \eta)\,] -  
d(\nu x \bar\nu)=\\
[0.2cm]
 -i{\bf\Xi_{\cal A}}d{\bf\bar\Xi^{\cal A}}=
[\,\nu^{\alpha}dl_{\alpha}+ \
\bar\nu^{\dot\alpha}d\bar l_{\dot\alpha} - 
4i({\bar\eta}_{m} d\eta^{m}- 
d{\bar\eta}^{m} \eta_{m})\,] -  d(\nu x \bar\nu), 
\end{array}
\end{equation}
where the first terms  in  r.h.s of  (\ref{3L})  
 are  dual  Wess-Zumino terms in 
the supertwistor and $\theta$-twistor spaces respectively
\footnote{ The Wess-Zumino  terms linear in derivatives were  
previously  considered in \cite {VZB}.}.  
 These  terms  are invariant under  the supersymmeties  
 (\ref{4'}) and (\ref{8'})  respectively  up to the total differential of the variation 
of the 
 scalar $(\nu x \bar\nu)$  which absorbs the space-time 
coordinates $x_{m}$.
 The  Wess-Zumino terms  in (\ref{3L}) 
may be used as the Lagrangians for two  dual supersymmetric actions of a particle with spin.
The  $\theta$-twistor representation of the action is given by the integral with respect to 
the proper time of  the particle 
\begin{equation}\label{LagZ}
S= \int  d\tau \{
[\,\nu^{\alpha}{\dot l}_{\alpha}+ {\bar\nu}^{\dot\alpha}\dot{\bar l}_{\dot\alpha} 
- 4i({\bar\eta}^{m}{\dot\eta}_{m}  - \dot{\bar\eta}_{m} \eta^{m}) \,]+ \lambda \,
s({\bf\Xi},{\bf\bar\Xi})\,\}, 
\end{equation}
where ${\dot f}=\frac{df}{d\tau}$ and $\lambda$ is the Lagrange multiplier
  fixing  the constraint (\ref{10'}) $s({\bf\Xi}, {\bf\bar\Xi})=0$.  
The action  (\ref{LagZ}) is invariant under the proper time 
reparametrizations and automatically introduces  the correct kinetic term for 
 the complex Ramond vector  encoding the spin degrees of  freedom
 of the massless spinning particle.

\section{Conclusion}

 Stimulated by the Gross-Wess observation \cite{GW} about the {\it
 spin} structures in scattering amplitudes of massless particles as the
 {\it obstructions} preventing the scale symmetry extension up to the
 conformal symmetry,  we have addressed the same question to generalized 
 superspaces with an  inherent chiral spin structure. On this way the
 supersymmetric twistors called the $\theta$-twistors and {\it dual}
 to the well known supertwistors \cite{Fbr}, \cite{Witt1} were
 revealed. The fermionic constituent of the $\theta$-twistor is
 presented by the composite grassmannian Ramond vector \cite{VZ} or by
 the chiral superspace coordinate $\theta_{\alpha}$ \cite{Z}
 contrarily to the scalar grassmannian constituent of the
 supertwistor. 
We established  that the super-Poincare and scale covariant chiral 
structures, associated with  the $\theta$-twistors, create some 
obstacles  for the (super)conformal boost realization. 
As a result,  the triplets creating  the 
 $\theta$-twistor superspace form representations of only  the maximal subgroup of the 
{\it superconformal} group. 
In the case when the $\theta$-twistor is realized  by three spinors 
the possibility  appears to
 restore half of  the (super)conformal boosts. The partial restoration goes 
 by means of mixing  
of the holomorphic and antiholomorphic triplets forming  the
 $\theta$-twistor space. It is interesting to understand a possible role
of this mixing in Yang-Mills theory for the description of scattering 
amplitudes different from the MHV amplitudes.

The $\theta$-twistor construction is  automatically generalized to the
case of extended supersymmetries just as the supertwistor
construction \cite{Fbr} and it is interesting to investigate the
geometrical properties of the corresponding supermanifolds
  along the line developed in \cite{GN}.

Because the Ramond vector  naturally appears
 as the $\theta$-twistor constituent it may  shed a new light on the mystery of 
the GSO projection \cite{GSO}.\footnote{Taking into account of the spontaneous vacuum
 transitions \cite{VZP} in the Veneziano and Neveu-Schwarz dual models
 has given an alternative mechanism of the tachyon elimination. 
In the Neveu-Schwarz  model this mechanism
 reveals the existence of the broken symmetry group with an infinite
 number of generators containing the group $ SU(2)\times SU(2)\times
 U(1)\times U(1)...\times U(1)...$ as a subgroup.} 
In this connection it is interesting to construct new supersymmetric
 actions of particles, strings and branes in the $\theta$-twistor space and 
to understand more on the connections between diffeomorphisms,
 ${\kappa}$-transformations and non-linear realizations \cite{W}, \cite{VZ}, \cite{STVZ},
 \cite{GKW}, \cite{Zemb}, \cite{BZ1}. We hope to study these issues elsewhere.

\section{Acknowledgements}

The author is grateful to Fysikum at the Stockholm University for kind
hospitality and Ingemar Bengtsson and Steven Giddings for useful
discussions. The work was partially supported by the grant of the
Royal Swedish Academy of Sciences.

\end{document}